\documentclass[conference]{IEEEtran}
\usepackage{tabularx}
\usepackage[T1]{fontenc}
\usepackage{algorithm}
\usepackage[noend]{algpseudocode}
\usepackage{graphicx}
\usepackage{amsmath,amssymb,amsfonts,stfloats}
\usepackage{array}
\usepackage{textcomp}
\usepackage[font=small]{caption}
\usepackage{paralist}
\usepackage{lettrine}
\usepackage{cite}
\usepackage{xcolor}
\usepackage{cancel}
\usepackage{booktabs}
\usepackage{mathtools}
\usepackage[caption=false,font=footnotesize]{subfig}
\usepackage{url}
\usepackage{acronym}
\usepackage{multicol}

\begin{document}

\title{Vehicular Blockage Modelling and Performance Analysis for mmWave V2V Communications\\
}
\author{Kai Dong, Marouan Mizmizi, Dario Tagliaferri and Umberto Spagnolini \\
\IEEEauthorblockA{Dipartimento di Elettronica, Informazione e Bioingegneria (DEIB), Politecnico di Milano, Milan, Italy. \\
Email: \{kai.dong, marouan.mizmizi, dario.tagliaferri, umberto.spagnolini\}@polimi.it}
}
\maketitle

\begin{abstract}
Vehicle-to-Everything (V2X) communications are revolutionizing the connectivity of transportation systems supporting safe and efficient road mobility. To meet the growing bandwidth eagerness of V2X services, millimeter-wave (e.g., 5G new radio over spectrum 26.50 - 48.20 GHz) and sub-THz (e.g., 120 GHz) frequencies are being investigated for the large available spectrum. Communication at these frequencies requires beam-type connectivity as a solution for the severe path loss attenuation. However, beams can be blocked, with negative consequences for communication reliability. Blockage prediction is necessary and challenging when the blocker is dynamic in high mobility scenarios such as Vehicle-to-Vehicle (V2V). This paper presents an analytical model to derive the unconditional probability of blockage in a highway multi-lane scenario. The proposed model accounts for the traffic density, the 3D dimensions of the vehicles, and the position of the antennas. Moreover, by setting the communication parameters and a target quality of service, it is possible to predict the signal-to-noise ratio distribution and the service probability, which can be used for resource scheduling. Exhaustive numerical results confirm the validity of the proposed model.
\end{abstract}


\section{Introduction}
Recent advances in Vehicle-to-Everything (V2X) communications accelerated the evolution of intelligent transportation systems towards a safe and efficient traffic management \cite{3GPP2018}. Currently deployed V2X technologies operates at sub-6 GHz frequencies \cite{coll2019sub} and can meet the requirements of basic V2X services due to the limited available bandwidth. Future 6G V2X is foreseen to support a wide range of services \cite{hakeem20205g} exploring high frequencies in the Millimeter Wave (mmWave) or even sub-THz bandwidths \cite{jameel2018propagation}. However, the wireless propagation at these frequencies is subject to severe path-loss attenuation. Collimated communication and beamforming technology is a viable solution to increase the communication range, but mmWave/sub-THz beams are subject to frequent blockage, which in vehicular scenarios such as Vehicle-to-Vehicle (V2V) links, represents the killing issue. According to the Third Generation Partnership Project (3GPP) \cite{3GPPTR38901}, there are four main propagation conditions: \textit{(i)} Line-of-Sight (LoS); \textit{(ii)} blockage due to a building (NLoSb); \textit{(iii)} blockage due to foliage (NLoSf) and \textit{(iv)} blockage due to vehicles (NLoSv). In this setting, blockage modeling and prediction are essential for the proper design of robust V2X systems to enable the critical requirements of ultra-high reliability and ultra-low latency.
 
Blockage impact on V2X communication has been receiving increasing attention in the recent literature. In \cite{boban2019multi}, V2V channel measurements in urban and highway roads considering different frequency bands in the presence of vehicle blockage were experimentally analyzed. At $30$ GHz, the blockage from a vehicle causes a power loss ranging from $10$ to $20$ dB \cite{park2017millimeter}, depending on the transmitting vehicle (TxV) and receiving vehicle (RxV) position, and it increases by $5.5-17$ dB when multiple vehicles are simultaneously blocking the LoS. Moreover, based on an extensive measurement campaign \cite{R1-1807672}, the 3GPP suggests modeling the vehicle blockage loss as a lognormal random variable added to the path loss, while in \cite{3GPPTR37885} the \textit{a-priori} probabilities of LoS and NLoSv are derived without considering the traffic density. The authors in \cite{boban2016modeling} extended the 3GPP model including the NLoSb condition and three traffic density levels, i.e., low, medium, and high. A further extension of the previous works is in \cite{giordani2019path}, considering the empirical blockage probabilities in highway and urban scenarios. However, the number of blocking vehicles and the variability of their shapes, particularly of the height, is not considered. The authors of \cite{eshteiwi2020impact} derived the end-to-end outage probability assuming the presence of a single blocking vehicle with a height modelled as a Gaussian distribution. However, the case of multiple vehicles simultaneously blocking the LoS has never been considered, including the contribution of the NLoSv to the communication link.

This paper addresses this gap in the literature by presenting an analytical model of SNR distribution and the service probability in highway V2V scenarios considering the impact of \textit{(i)} multiple blocking vehicles, \textit{(ii)} random vehicles' height, and \textit{(iii)} traffic density. The path-loss and blockage models are based on the 3GPP recommendations, and the vehicular traffic is obtained by extending the work in \cite{abul2007modeling, cui2018vehicle}, where the traffic distribution is represented by a Poisson Point Process (PPP). This analytical tool can be useful in designing solutions for reliable and robust V2X networks and for resource scheduling to mitigate the blockage effects. The presented models are validated through extensive numerical simulations with vehicular traffic generator under various traffic conditions.

The remainder of this paper is organized as follows: Section \ref{SystemModel} presents the system model and SNR derivation. The proposed vehicle blockage is derived in Section \ref{BlockageModel}, while Section \ref{NumericalSimulation} shows the numerical validation. Finally, closing remarks are discussed in Section \ref{Conclusion}.

\section{System Model} \label{SystemModel}

\begin{figure}[t!]
	\centering
	\includegraphics[width=0.9\columnwidth]{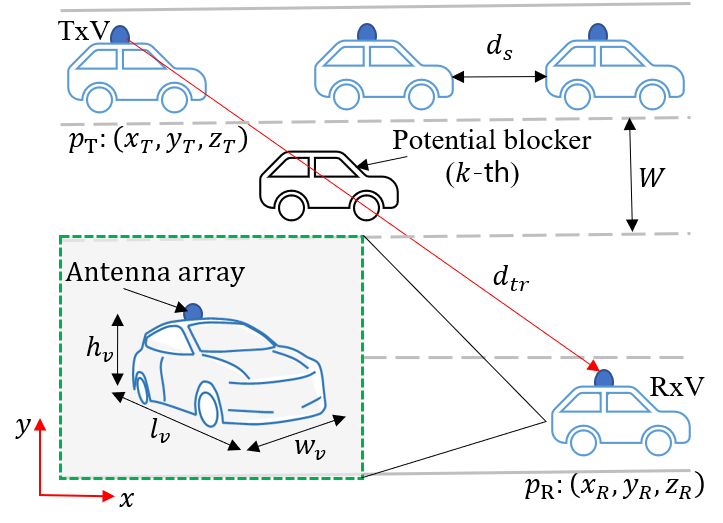}
     \caption{A linear highway scenario for sidelink V2V communications.}
    \label{Scenario}
\end{figure}

The considered scenario is a multi-lane highway road segment, depicted in Fig. \ref{Scenario}, where vehicles are randomly distributed with density $\rho$ [veh/m]. Vehicles are equipped with a mmWave/sub-THz V2V communication interface mounted either on the rooftop or on the frontal/rear bumper \cite{3GPPTR37885}, and are assumed to have a box-shaped occupation of size $w_v \times l_v \times h_v$, as shown in Fig. \ref{Scenario}.

For an arbitrary V2V link between a transmitting vehicle (TxV) and a receiving vehicle (RxV), the propagation path-loss (in dB) is \cite{3GPPTR37885}
\begin{equation} \label{PL-los}
\begin{split}
   PL(k) & = \underbrace{32.4 + 20\log_{10}(d_{tr}) + 20\log_{10}(f_c)}_{\mu_{\text{LoS}}} + A(k) + \chi
\end{split}
\end{equation}
where $d_{tr}$ is the distance between the TxV and RxV, $f_c$ is the carrier frequency (in GHz), $\chi \sim \mathcal{N}(0, \sigma_{sh}^2)$ represents the lognormal distributed shadowing component, and $A(k) \sim \mathcal{N}(\mu(k), \sigma^2(k))$ accounts for an additional attenuation due to blockage from $k$ vehicles. 
%
%
Assuming independence between the shadowing component and the blockage component, we can write the path-loss as
\begin{equation}\label{eq:tot_pl}
    PL(k) \sim
    \mathcal{N}\left(\underbrace{\mu_{\text{LoS}}+\mu(k)}_{\mu_{PL}(k)}, \underbrace{\sigma_{sh}^2 + \sigma^2(k)}_{\sigma_{PL}^2(k)}\right)
\end{equation}
where $\mu_{\text{LoS}}$ is the deterministic term in \eqref{PL-los}. 

The signal-to-noise ratio (SNR) $\gamma(k)$ (in dB) is
\begin{equation}
    \gamma(k) = P_{t} + G_{t} + G_{r} - PL(k) - P_n
\end{equation}
where $P_{t}$ is the transmitted power, $G_{t}$ and $G_{r}$ are the TxV and RxV array gains, respectively, $P_n$ is the noise power.  
The SNR distribution for distance $d_{tr}$ and for $k$ vehicle blockers at the LoS V2V link is
\begin{equation}\label{eq:snr_dist}
    f_{\gamma^{(k)}}(\gamma|d_{tr}) = \frac{1}{\sqrt{2\pi\sigma^2_{PL}(k)}} \exp\left\{-\frac{(\gamma - \mu_{\gamma}(k))^2}{2\,\sigma_{PL}^2(k)}\right\}
\end{equation}
with mean value $\mu_\gamma(k) = P_{t} + G_{t} + G_{r} - \mu_{PL}(k) - P_n$ and variance $\sigma_{PL}^2(k)$, both dependent on the propagation condition between TxV and RxV.

\section{Vehicle Blockage Modelling and Performance Evaluation}\label{BlockageModel}
\begin{figure}[b!]  
	\centering
	\includegraphics[width=0.5\textwidth]{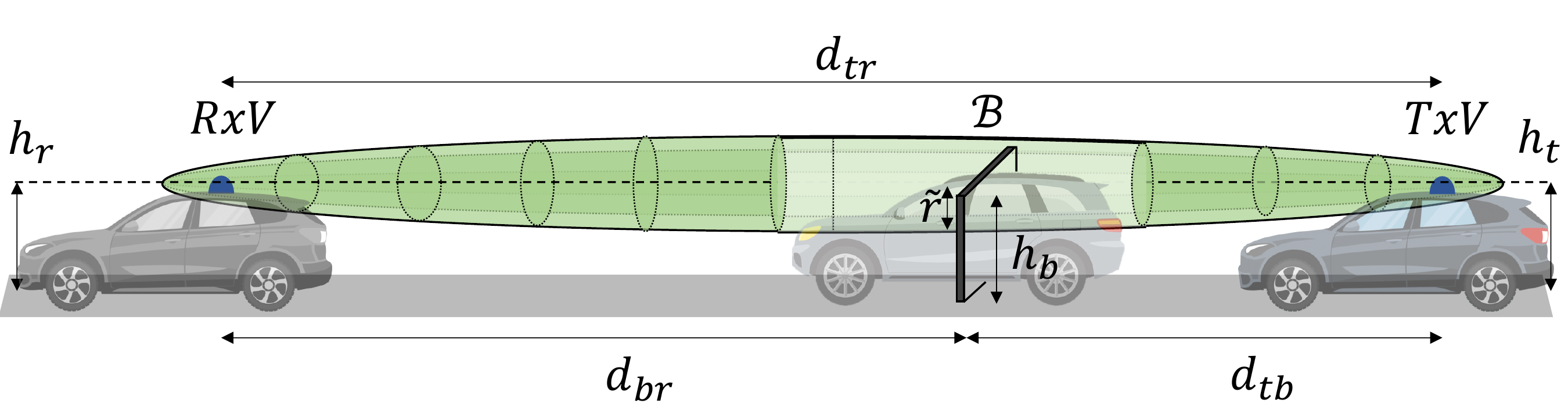}
    \caption{First Fresnel Ellipsoid blockage characterization considering the height of the vehicles}
    \label{FresnellEll}
\end{figure}
From the electromagnetic perspective, the LoS propagation is affected when one (or more) blocking vehicle obstructs the first Fresnel ellipsoid. 
Let's assume the height of each vehicle as a Gaussian random variable $h_v \sim \mathcal{N}(\mu_{v}, \sigma_{v}^2)$ \cite{akhtar2014vehicle} and a single blocker $\mathcal{B}$ at distance $d_{tb}$ from TxV and $d_{br}$ from RxV (along the LoS line), as depicted in Figure \ref{FresnellEll}. We can use the knife edge theory on link propagation~\cite{SavazziSpagnolini} to express the effective height of the first Fresnel ellipsoid at the blocker position as
\begin{equation}\label{eq:h_eff}
    \tilde{h} = (h_r - h_t) \frac{d_{tb}}{d_{tr}} + h_t - 0.6\, \tilde{r}
\end{equation}
where $h_t$ and $h_r$ are the heights of the TxV and RxV, respectively and $\tilde{r}$ is the radius of the first Fresnel ellipsoid at relative distance $d_{tb}$ and $d_{br}$
\begin{equation}\label{eq:rf}
    \tilde{r} = \sqrt{\lambda_c \frac{d_{tb} d_{br}}{d_{tb} + d_{br}}}.
\end{equation}
with $\lambda_c$ being the carrier wavelength. For the propagation at mmWave frequencies ($30-300$ GHz), the term $\tilde{r}$ might be meaningful in \eqref{eq:h_eff}. For a V2V system operating in 5G NR FR2 bands ($28$ GHz) and a maximum TxV-RxV distance $d_{tr} = 200$ m, $\tilde{r}\approx 1$ m and therefore must be taken into account. Differently, for sub-THz systems ($>100$ GHz) and shorter distances ($<50$ m), $\tilde{r}$ can be reasonably neglected.
Hence, the Fresnel ellipsoid height is, in general, 
\begin{equation}\label{eq:h_eff_app}
    \tilde{h} = h_r \frac{d_{tb}}{d_{tr}} + h_t \frac{d_{br}}{d_{tr}} - 0.6\, \tilde{r}\sim \mathcal{N}\left(\tilde{\mu}, \tilde{\sigma}^2\right)
\end{equation}
with mean $\tilde{\mu} = \mu_{v} - 0.6\, \tilde{r}$ and variance $\tilde{\sigma}^2 = \sigma_{v}^2$, assuming the independence between the TxV and RxV heights $h_t$ and $h_r$. Thus, a blockage occurs when the random blocker height $h_b\sim \mathcal{N}(\mu_{b}, \sigma_{b}^2)$ is larger than $\tilde{h}$, i.e., $h_{eff} = h_b - \tilde{h} > 0$. The probability of having a blockage given the presence of a vehicle $\mathcal{B}$ between the TxV and RxV can be computed as:
\begin{equation}\label{Problockage}
    \mathbb{P}\left(\text{NLoSv} | d_{tr},\mathcal{B} \right) = Q\left(\frac{h_{eff} - \mu_{eff}}{\sigma_{eff}}\right)
\end{equation}
where $\mu_{eff} = \mu_{b} - \tilde{\mu}$, $\sigma_{eff}^2 = \sigma^2_{b} + \tilde{\sigma}^2$, and $Q(x)$ is the Gaussian Q-function.

The probability of having $k \geq 0$ blockers depends on the distribution of vehicles in the scenario and on the distance $d_{tr}$ between TxV and RxV. To ease the analytical derivation, we first determine the probability of having $k$ blockers within a given distance $d_{tr}$, and subsequently, we will remove this constraint obtaining the unconstrained SNR distribution.

In a $M$-lane highway V2V scenario, the probability of having a vehicle on any arbitrary lane is $1/M$. Therefore, the probability of having both TxV and RxV on the same lane is $M(1/M^2)=1/M$, while the probability of having them on different lanes is $1-(1/M)$. In the two cases, the maximum number of $k$ blockers is different.

\begin{figure}[t!]  
	\centering
	\subfloat[Same lane]{\includegraphics[width=0.5\textwidth]{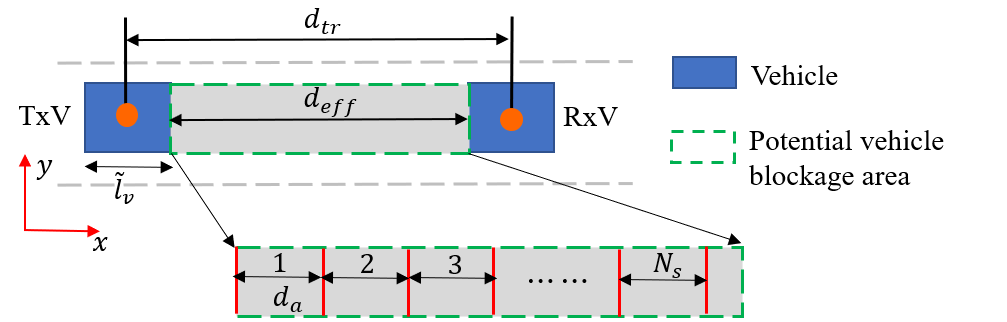}} \\
    \subfloat[Different lanes] {\includegraphics[width=0.5\textwidth]{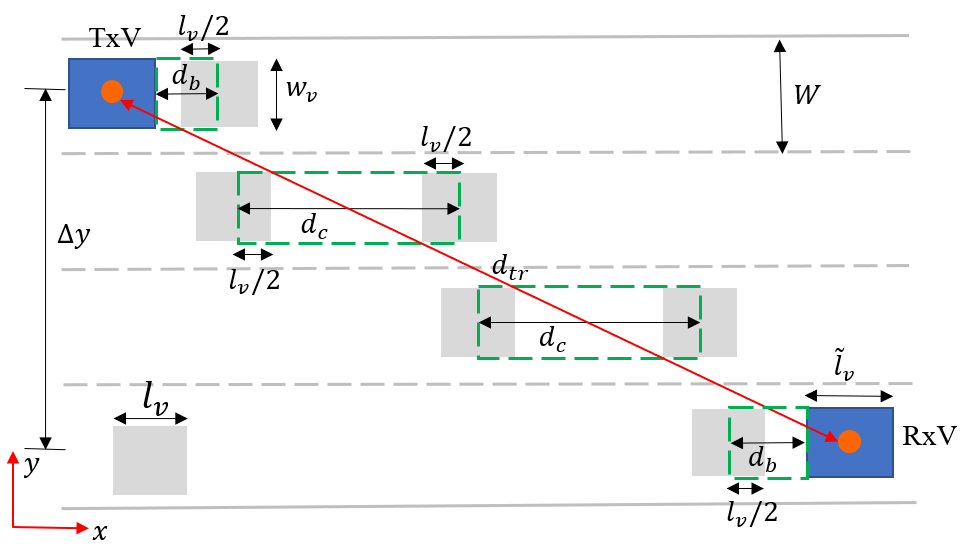}}
    \caption{Proposed vehicle blockage modelling approaches: (a) both TxV and RxV locate at the same lane; (b) TxV and RxV in different lanes.}
    \label{BlockageModelling}
\end{figure}

\subsection{TxV and RxV on the Same Lane}

When the TxV and RxV are on the same lane, as shown in Figure \ref{BlockageModelling}a, the maximum number of blockers, for example in a traffic jam, is $N_s = \lfloor d_{eff} / d_a\rfloor$, where the size of the vehicle occupancy is
\begin{equation}
    d_a = l_v + d_s,
\end{equation}
that is the effective length of the vehicle incorporating a safety distance $d_s$, while $d_{eff} = d_{tr} - l_v$ is the effective distance between TxV and RxV. Dividing the lane into $N_s$ vehicle occupancy slots of size $d_a$, the probability of having a simultaneous blockage by $k$ vehicles is thus given by the probability of having $k$ occupied slots out of $N_s$, that is given by the Bernoulli probability
\begin{equation} \label{eq:pro-B-SL}
    \mathbb{P}_{SL}\left(\text{NLoSv}^{(k)}|d_{tr}\right) =  
    \begin{pmatrix}
     N_s \\
     k
    \end{pmatrix}
    \mathcal{P}_{a}^{k}(1-\mathcal{P}_{a})^{N_s-k}
\end{equation}
where $\mathcal{P}_{a}$ is the probability that a single slot is occupied by a blocker. This latter term is
\begin{equation} \label{eq:p_block}
\begin{split}
    \mathcal{P}_{a} = \mathbb{P}\left(\text{NLoSv} | d_a,  \mathcal{B} \right) \, \mathbb{P}\left(\mathcal{B} \right) 
    = Q\left(\frac{h_{eff} - \mu_{eff}}{\sigma_{eff}}\right) \Gamma \, e^{-\Gamma}, 
\end{split}
\end{equation}
assuming that vehicles on a lane are distributed according to a linear Point Poisson Process (PPP) as in \cite{abul2007modeling}\cite{cui2018vehicle}, with $\Gamma = \rho\,d_a$.

\subsection{TxV and RxV on Different Lanes}

Assuming the TxV and RxV are on two distinct lanes, the blocking vehicles can occupy the same lane as the TxV and/or RxV or different lanes for $M>2$, as shown in Figure \ref{BlockageModelling}b. Similar to the previous setup, we divide the area between TxV and RxV into $\widetilde{M}$ possible occupancy slots (where $\widetilde{M}$ denotes the amount of spanned lanes by the V2V link) that, if occupied by a vehicle, could block the LoS. The size of the occupancy slots determines the blockage probability, and we can identify two types, i.e., blockage from the same lane of the TxV or RxV and different lanes. The sizes can be derived geometrically, that for the first type is
\begin{equation}\label{eq:d_b}
    d_{b} = \frac{w_v \sqrt{d_{tr}^2 - \Delta y^2}}{2\Delta y}
\end{equation}
and for the second type is
\begin{equation}\label{eq:d_c}
    d_{c} = \frac{w_v\sqrt{d_{tr}^2 - \Delta y^2 }}{\Delta y} + l_v
\end{equation}
where $\Delta y = n\,W$ is the lateral displacement among the lanes. The probability of having a blockage in occupancy slot type $b$ and $c$, namely $\mathcal{P}_b$ and $\mathcal{P}_c$, respectively, can be computed as in \eqref{eq:p_block}, with $\Gamma = \rho d_b$ and $\Gamma = \rho d_c$.
Since $d_b$ and $d_c$ depend on $n\,W$ for $n > 1$ (for $n = 0$, the TxV and RxV are on the same lane), we need to consider all possible displacements separately, i.e., $n = 1, ..., M-1$. For $n = 1$, the maximum number of blocking slots is $\widetilde{M}=2$ with an occupancy slot size computed as in \eqref{eq:d_b}. The probability of having $k$ blockers out of $2$ can be computed using the Bernoulli probability \eqref{eq:pro-B-SL}. For $n > 1$, the maximum number of blocking slots is $\widetilde{M}=n+1$. In this case, we have two slots' types with sizes computed as in \eqref{eq:d_b} and \eqref{eq:d_c}. The probability of having $k$ blockers out of $n+1$ is obtained by Poisson Binomial 
\begin{equation}
    \mathbb{P}(K=k) = \sum_{\mathcal{A}\in \mathcal{Q}_k} \prod_{i \in \mathcal{A}}\mathcal{P}_i \prod_{j\in \mathcal{A}^c} (1-\mathcal{P}_j)
\end{equation}
where $\mathcal{Q}_k$ is the set of all $k$ slots that can be selected from $\{1,\dots,n+1\}$ and $\mathcal{A}^c$ is the complement of $\mathcal{A}$, i.e., $\mathcal{A}^c = \{1,\dots,n+1\}\not\in \mathcal{A}$. For example, when $n = 2$ and $k = 2$, then $\mathcal{Q}_2 = \{\{1,2\},\{2,3\},\{1,3\}\}$. The number of combination in $\mathcal{Q}_k$ is given by the binomial coefficient
\begin{equation}
    \begin{pmatrix}
     n+1 \\
     k
    \end{pmatrix} = \frac{(n+1)!}{(n-k+1)!k!}
\end{equation}
To combine all the cases for different $n$, we need to compute the probability of having TxV and RxV with lateral displacement $n\,W$, that is
\begin{equation}
    \mathbb{P}\left(\Delta y = n\,W\right) = 2 \frac{M-n}{M^2}
\end{equation}
assuming that the probability of having a vehicle on an arbitrary lane is $1/M$.\\
\textit{Remark:} by considering all possible combination of $n=\{1,\dots,M-1\}$, the aggregated can be written as
\begin{equation}
    \frac{2}{M^2} + \sum_{n=1}^{M-1} n = \frac{M-1}{M}   
\end{equation}
knowing that $\sum_{i=1}^{m} i = m(m+1)/2$. By considering the same lane setup, i.e., $n=0$, the sum of the probabilities of all events equals 1.

The probability of having $k$ simultaneous vehicle blockers when TxV and RxV are on different lanes is derived in \eqref{eq:P_dl}, while the overall blockage probability constrained on having TxV and RxV at a distance $d_{tr}$ is in \eqref{eq:p_dtr}.
\begin{figure*}[t!]
    \centering
    \begin{equation}\label{eq:P_dl}
        \mathbb{P}_{DL}\left(\text{NLoSv}^{(k)}|d_{tr}\right) = \frac{2(M-1)}{M^2} 
        \begin{pmatrix}
             n+1 \\
             k
        \end{pmatrix} \mathcal{P}_b^{k} (1-\mathcal{P}_b)^{n+1-k}  + \sum_{n=2}^{M-1} \frac{2(M-n)}{M^2} \sum_{\mathcal{A}\in \mathcal{Q}_k} \prod_{i \in \mathcal{A}}\mathcal{P}_i \prod_{j\in \mathcal{A}^c} (1-\mathcal{P}_j) 
    \end{equation}
    
    \begin{equation}\label{eq:p_dtr}
        \mathbb{P}\left(\text{NLoSv}^{(k)}|d_{tr}\right) = \mathbb{P}_{DL}\left(\text{NLoSv}^{(k)}|d_{tr}\right) + \frac{1}{M} \mathbb{P}_{SL}\left(\text{NLoSv}^{(k)}|d_{tr}\right)
    \end{equation}
\end{figure*}

%

To obtain the unconstrained probability of blockage, the probability density function $f_D(d)$ of the distance $d_{tr} = \sqrt{(x_r-x_t)^2 + (y_r-y_t)^2}$ is required. $f_D(d)$ has been widely studied in geometric probability \cite{GeoProb} and can be derived as in \cite{10.2307/3212475}, assuming that $x_r \sim x_t \sim \mathcal{U}\{0, D\}$ with $D$ the maximum length of the considered highway scenario, and that $y_t$ and $y_r$ are discrete uniform random variables with $p(y) = 1/M$.

\subsection{Signal-to-Noise Ratio Distribution}

The SNR probability density function $f_{\gamma^{(k)}}(\gamma|d_{tr})$ in \eqref{eq:snr_dist} is Gaussian distributed with $\mu_\gamma(k)$ and $\sigma_\gamma^2(k)$ that depend on the number of blocking vehicles $k$. The unconstrained SNR distribution results in a Gaussian mixture, representing the different propagation conditions:
\begin{equation}\label{eq:snr_unconstrained}
\begin{split}
    f_\gamma(\gamma|d_{tr}) = & \,\mathbb{P}(\text{LoS}|d_{tr}) \, f_{\gamma^{(0)}}(\gamma|d_{tr}) \, +\\ 
    & + \sum_{k=1}^{B} \mathbb{P}(\text{NLoSv}^{(k)}|d_{tr})\, f_{\gamma^{(k)}}(\gamma|d_{tr})
\end{split}
\end{equation}
where $\mathbb{P}(\text{LoS}|d_{tr}) = 1 - \sum_k \mathbb{P}(\text{NLoSv}^{(k)}|d_{tr})$ is the LoS probability, and $B = \mathrm{max}(N_s,M)$ is the maximum possible number of blockers.

\section{Numerical Simulations}\label{NumericalSimulation}

\begin{figure}[b!]  
	\centering
	\includegraphics[width=0.5\textwidth]{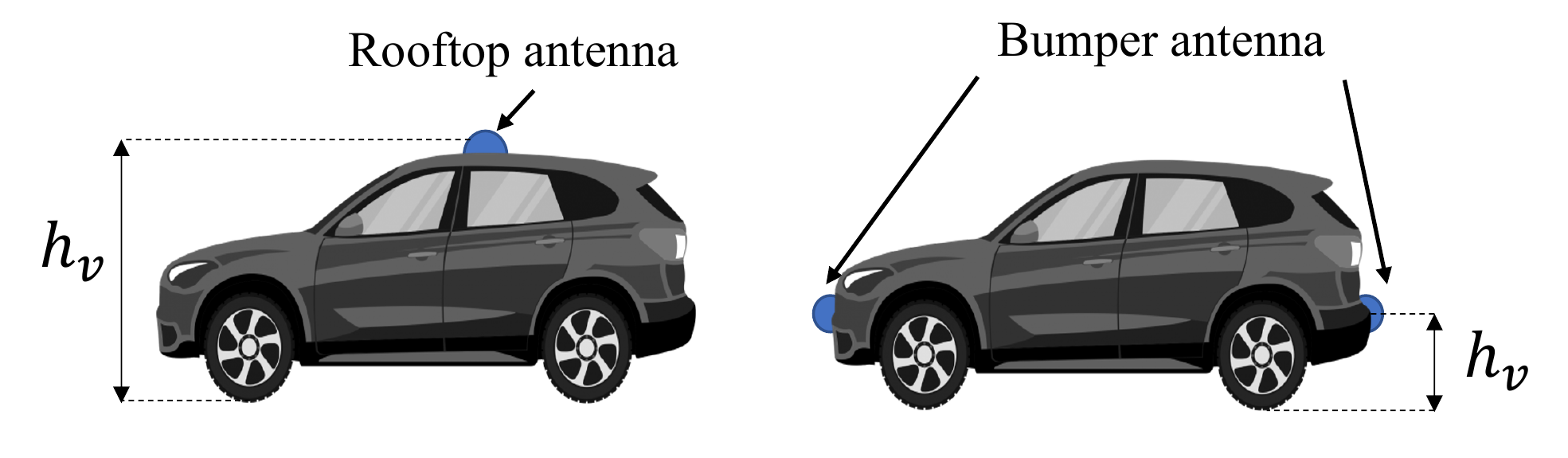}
    \caption{Rooftop vs bumper antenna configuration according to 3GPP \cite{3GPPTR37885}}
    \label{AntennaConf}
\end{figure}

This section presents the numerical validation of the proposed analytical framework. In particular, we derive the SNR distribution in \eqref{eq:snr_unconstrained} considering different scenario setups, and we analyze the communication performances in terms of service probability:
\begin{equation} \label{eq:Service-pro}
\begin{split}
    P_S = \mathbb{P}(\gamma \geq \gamma_{th}) &= \mathbb{P}(\text{LoS}|d_{tr})\, F_{\gamma^{(0)}}(\gamma_{th}) + \\
    & + \sum_{k=1}^{B} \mathbb{P}(\text{NLoSv}^{(k)}|d_{tr}) \, F_{\gamma^{(k)}}(\gamma_{th}),
\end{split}
\end{equation}
where $F_{\gamma^{(k)}}(\gamma_{th})$ is the normal cumulative distribution function. Unless otherwise specified, the simulation parameters in Table \ref{ParameterSetups} are used. These parameters are selected based on the 3GPP recommendation in \cite{3GPPTR37885}, and similarly for the antenna configuration, as depicted in Figure \ref{AntennaConf}.


%

\begin{table}[t!]
\caption{Parameter Setups}
\begin{center}
\begin{tabular}{l|c|c}
\toprule
\textbf{Parameter} & \textbf{Symbol} & \textbf{Value} \\
\toprule
Transmit power & $P_t$ &  $0$ dBm\\
Transmit antenna gain & $G_t$ &  $10$ dB \\
Receive antenna gain & $G_r$ &  $10$ dB \\
Noise Power & $P_n$ &  $-85$ dBm \\
Vehicle length, width & $[l_v, w_v]$ &  $[5, 1.8]$ m \\
Vehicle height mean and std. dev. & $\mu_{h,v}, \sigma_{h,v}$ & $[1.5, 0.08]$ m\\
Min. inter-vehicle safety distance & $d_s$ &  $2.5$ m \\
Scenario length & $D$ &  $200$ m \\
Number of lanes & $M$ &  $3$ \\
\bottomrule
\end{tabular}
\label{ParameterSetups}
\end{center}
\end{table}

\begin{figure}[t!]  
	\centering
	\subfloat[]{\includegraphics[width=0.9\columnwidth]{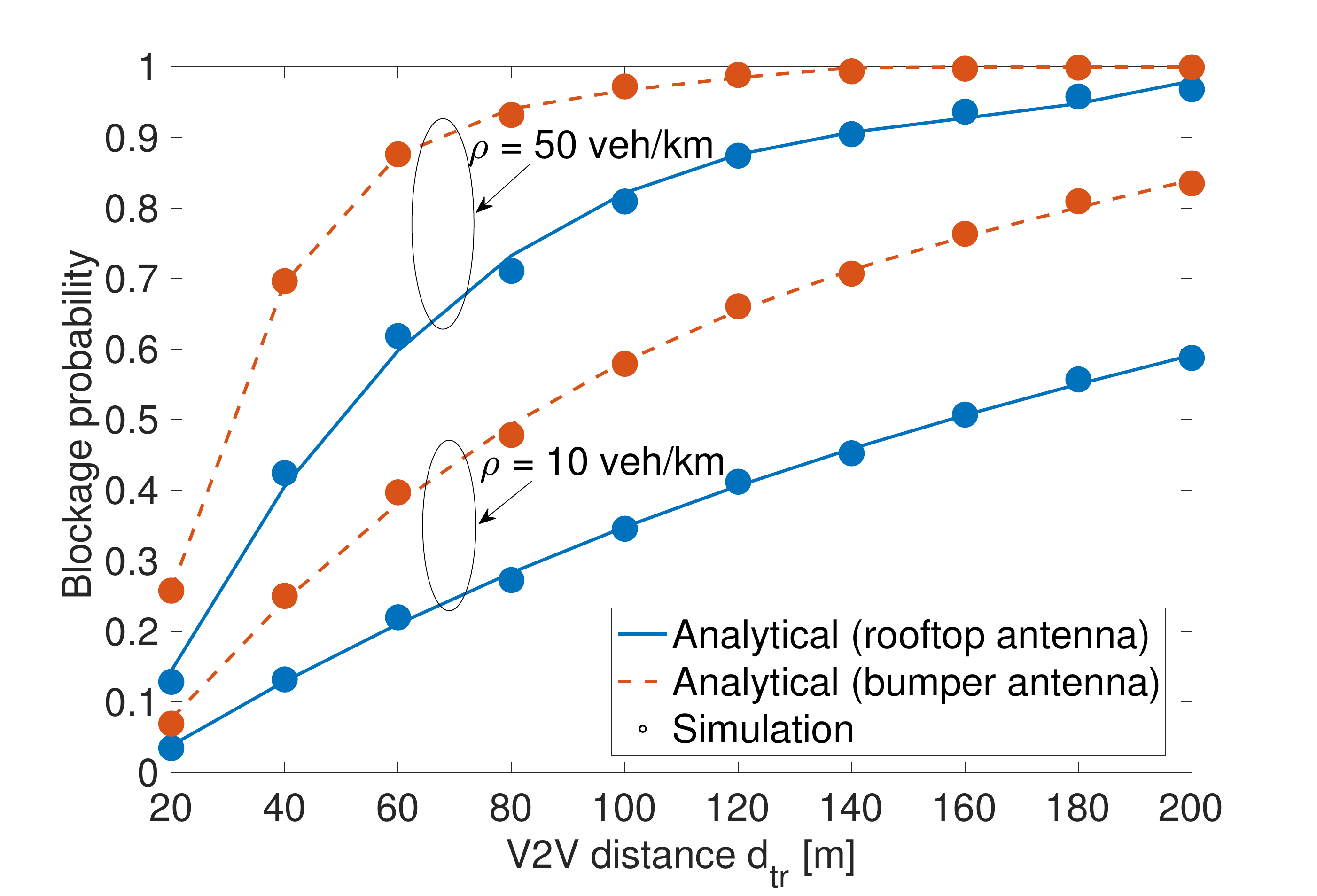}} \\
    \subfloat[]{\includegraphics[width=0.9\columnwidth]{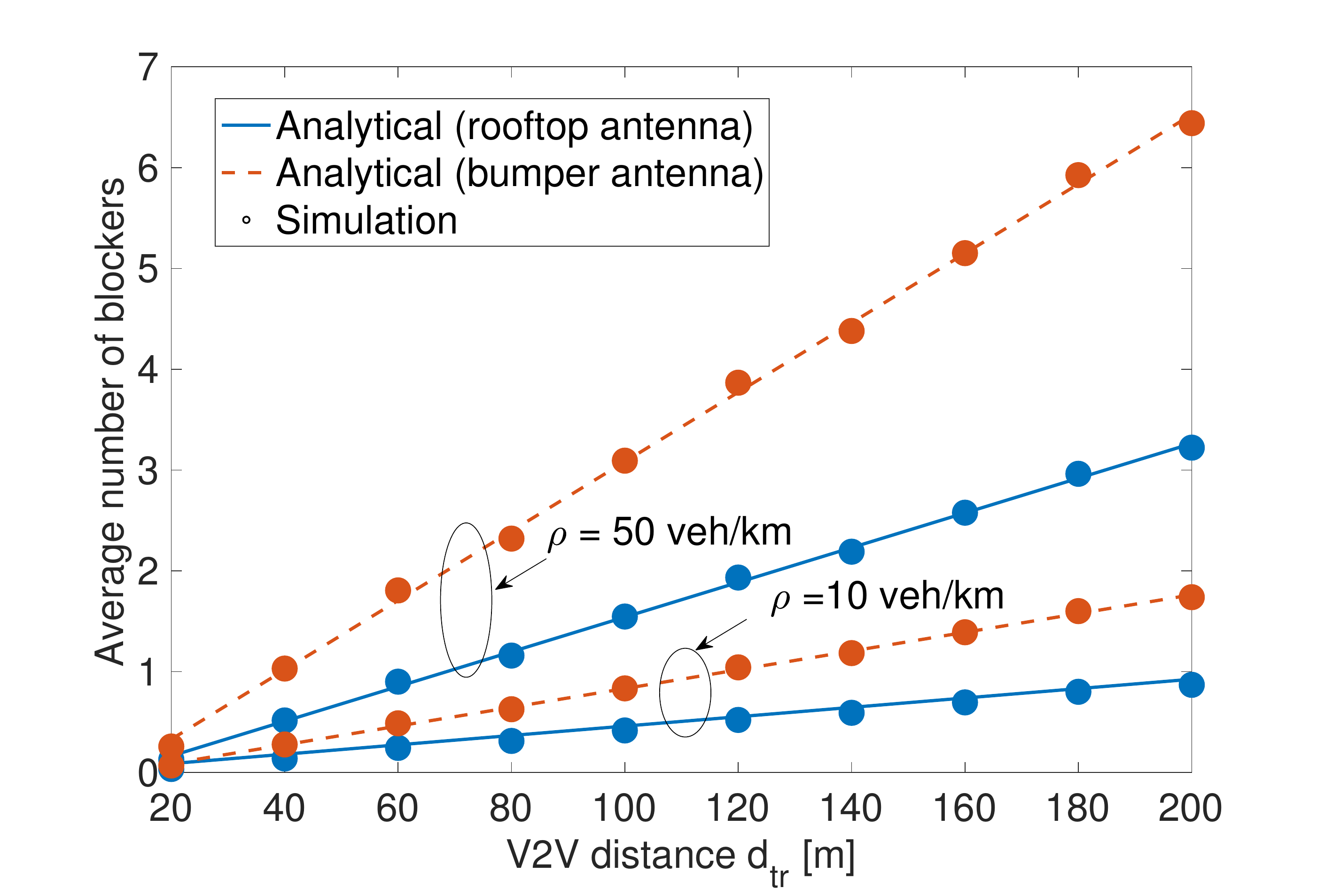}}
    \caption{Blockage probability and average number of vehicle blockers varying the TxV and RxV inter-distance $d_{tr}$: (a) Blockage probability; (b) average number of blockers.}
    \label{fig:blockProb}
\end{figure}

\begin{figure}[t!]  
	\centering
	\subfloat[]{\includegraphics[width=0.9\columnwidth]{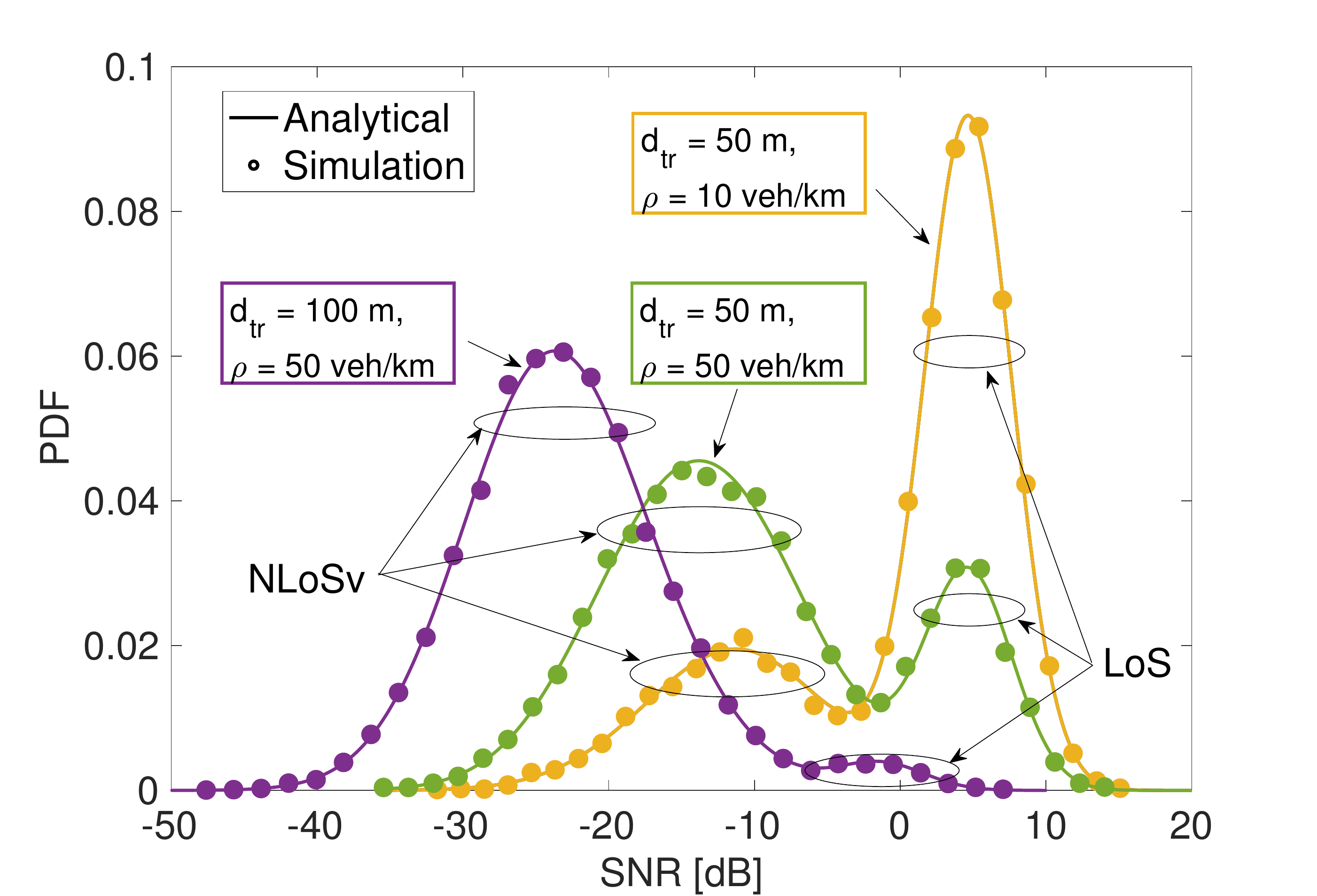}}\\
    \subfloat[]{\includegraphics[width=0.9\columnwidth]{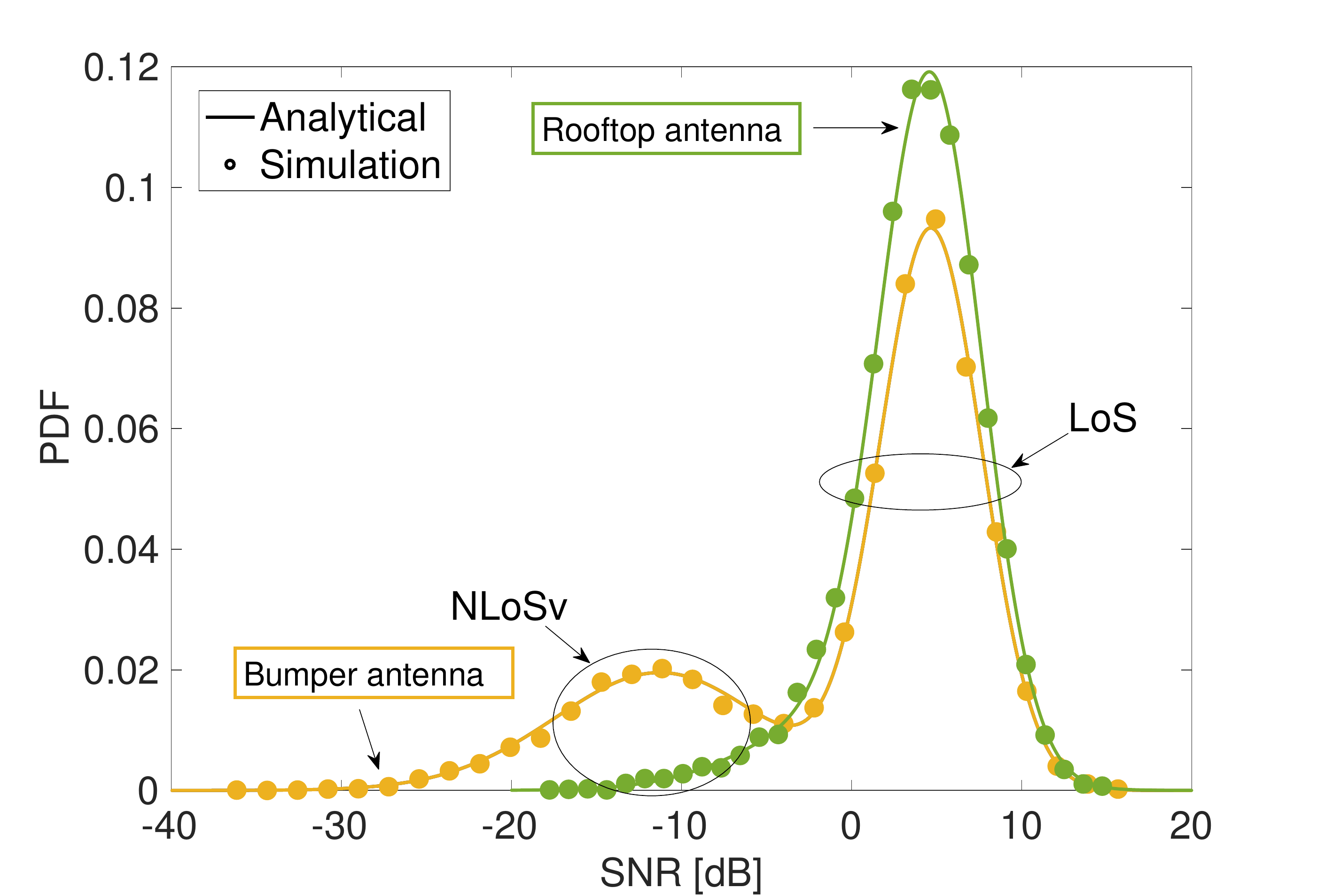}}
    \caption{SNR distribution fitted with Monte Carlo simulations, conditioned to distance $d_{tr}$ for both TxV and RxV in the same lane: (a) different distance $d_{tr}$ and density $\rho$, and 
    (b) different antenna configurations (i.e., rooftop and bumper antenna setups).}
    \label{fig:snr_dist}
\end{figure}

Exhaustive Monte Carlo simulations are carried out to validate the proposed analytical model. Figure \ref{fig:blockProb} (a) depicts the blockage probability varying the distance $d_{tr}$ between TxV and RxV and considering two traffic densities, namely $\rho = 10$, and $50$ veh/Km. This result shows that for high distances, the blockage probability attains around $50$\% and $90$\% in low and high traffic densities, respectively. This can also be observed in Figure  \ref{fig:blockProb} (b), where, for the same scenario, the average number of blockers is reported. The proposed model is used to predict the communication performances by deriving the SNR distribution in \eqref{eq:snr_dist} and the service probability in \eqref{eq:Service-pro}. Figure \ref{fig:snr_dist} shows the predicted SNR distribution under various traffic densities and distances. In particular, for the considered setting, we observe that the average SNR in Figure \ref{fig:snr_dist} (a) is below $0$ dB for high traffic density due to the higher blockage probability. Figure \ref{fig:snr_dist} (b) depicts the predicted SNR distribution when considering the antenna mounted on the rooftop and the bumper in the same scenario. As expected, the antennas on the bumper experience a loss of approximately $4.2$ dB in SNR. This result suggests considering rooftop antenna design for better communication performances as it shows more robustness against blockage. 

\begin{figure}[t!]
	\centering
	\includegraphics[width=\columnwidth]{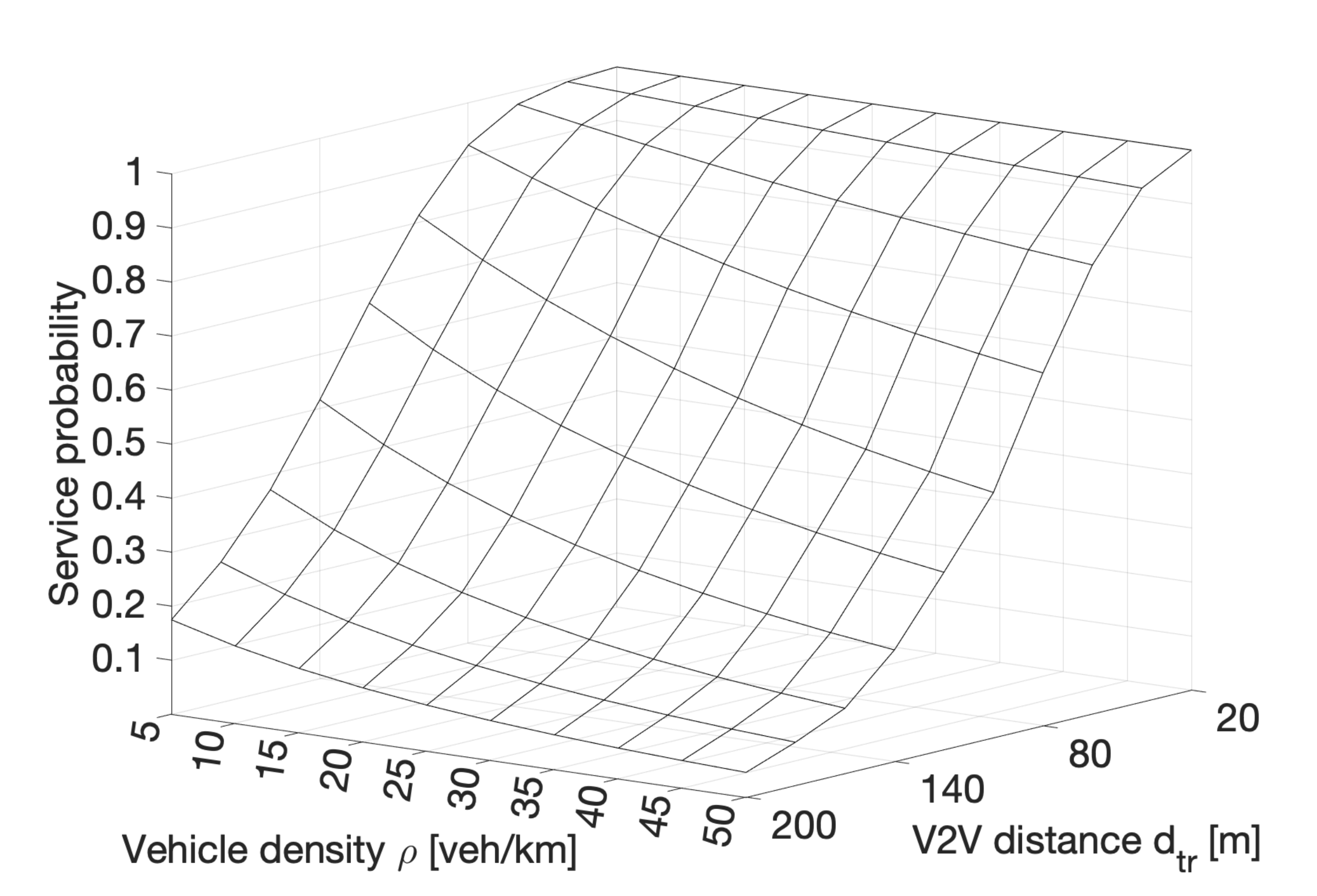}
     \caption{Service probability constrained by a inter-vehicle distance $d_{tr}$ under different traffic densities based on the derived SNR in \eqref{eq:snr_unconstrained} constrained by $\gamma_{th} = 0$ dB.}
    \label{fig: sp_density}
\end{figure}

\begin{figure}[!htbp]  
	\centering
	\subfloat[]{\includegraphics[width=0.9\columnwidth]{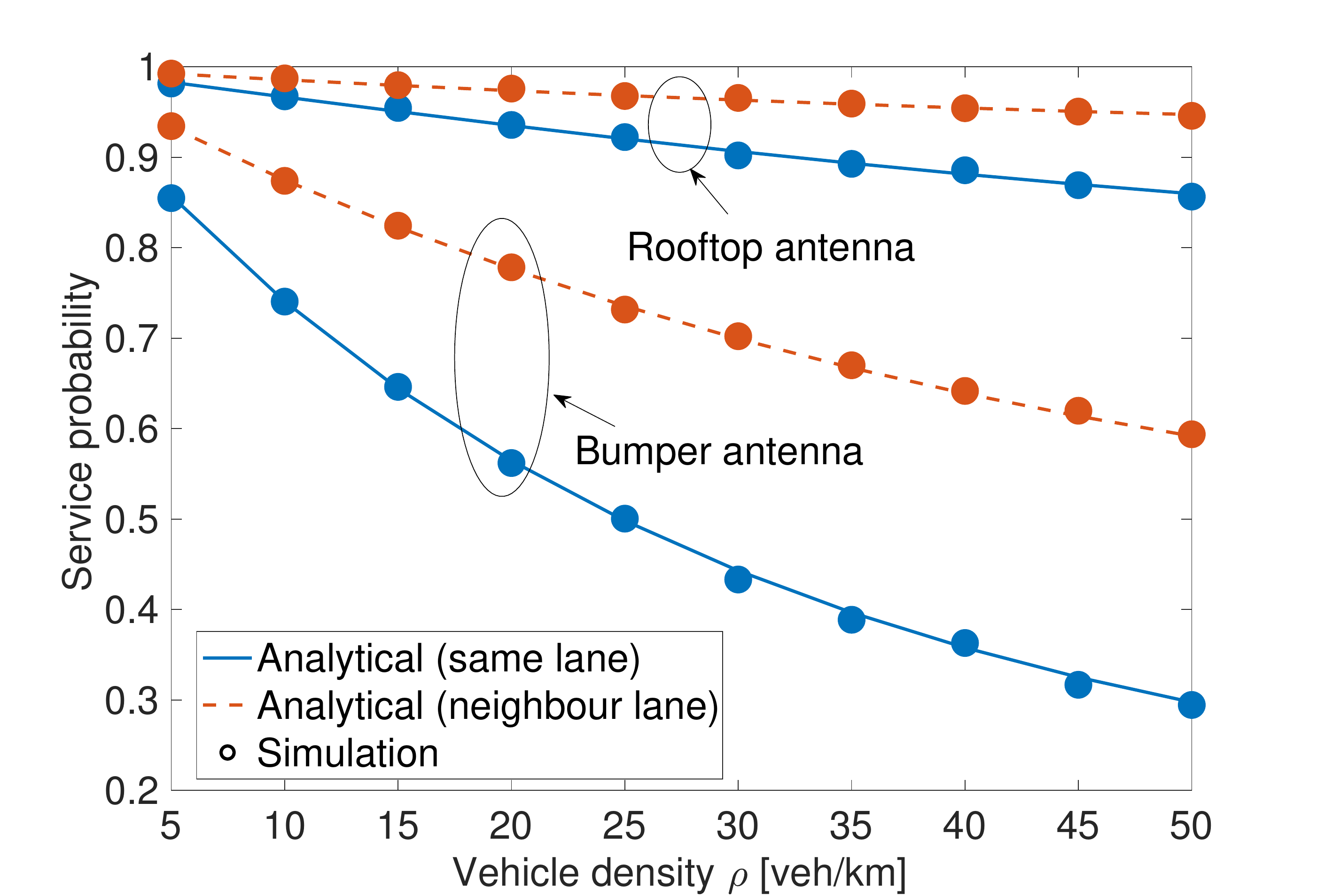}} \\
    \subfloat[]{\includegraphics[width=0.9\columnwidth]{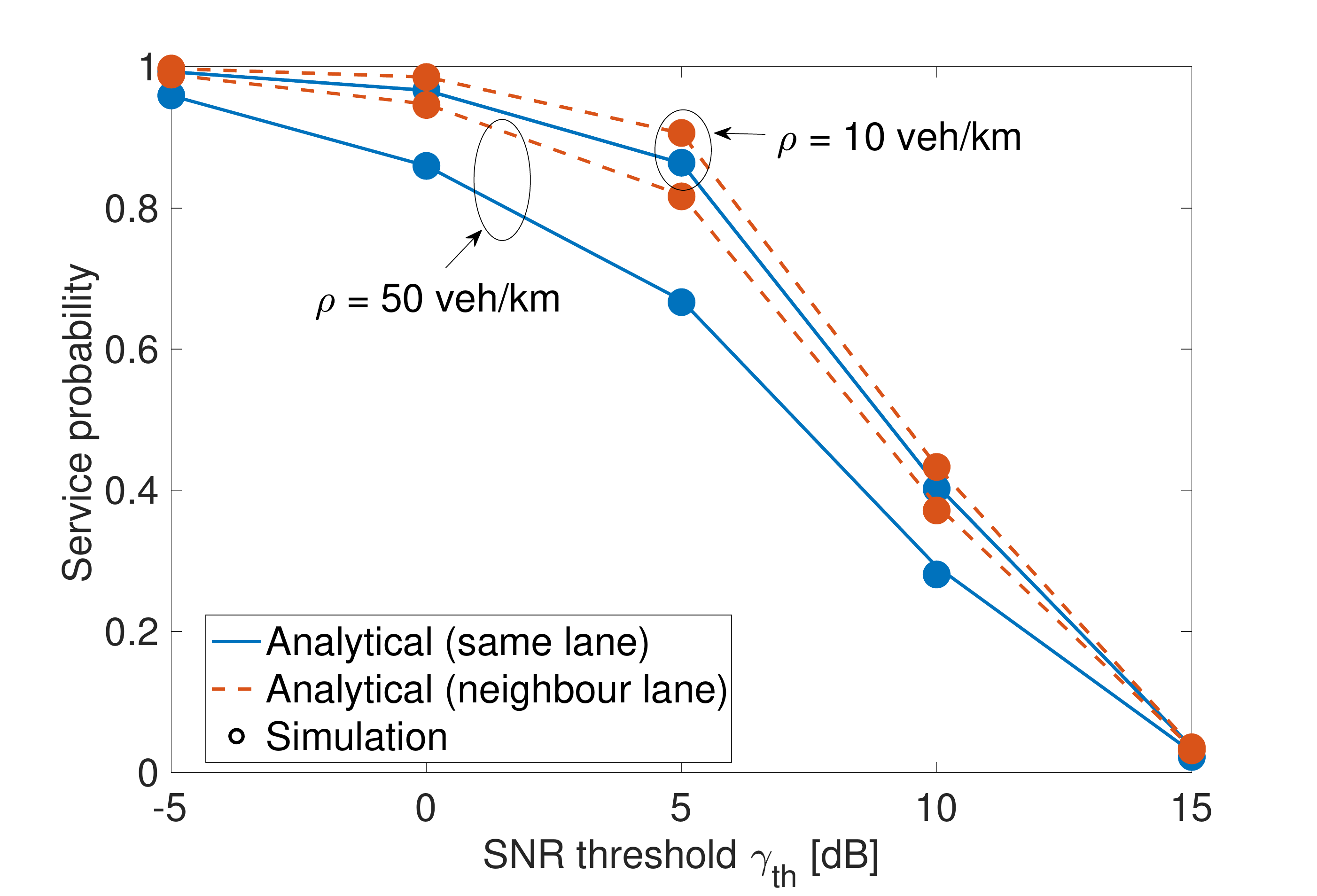}}
    \caption{Service probability evaluation ($d_{tr} = 50$ m) varying with (a) density $\rho$ for cases of TxV and RxV in the same lane and neighbour lanes ($\gamma_{th} = 0$ dB); (b) SNR threshold $\gamma_{\text{th}}$ constrained by different service categories.}
    \label{fig:SP_SNRth}
\end{figure}

The predicted SNR distribution can be used to derive the service probability using \eqref{eq:Service-pro} for a given QoS, i.e., SNR threshold $\gamma_{th}$. Figure \ref{fig: sp_density} depicts the service probability varying both traffic density rho and distance $d_{tr}$ for an SNR threshold $\gamma_{th} = 0$ dB. Even at low density, we have a poor service probability; therefore, direct V2V communication can support data rate demanding applications only for a limited range. This trend is more severe when the antennas are mounted on the bumper, as shown in Figure \ref{fig:SP_SNRth} (a), and when increasing the QoS, as shown in Figure 6 (b). 

The analytical tools presented in this paper can be valuable in designing solutions that increase the reliability and robustness of future V2X networks, for example, scheduling resources for relaying from other vehicles or roadside units to mitigate the blockage effect.

%

%
\section{Conclusion}\label{Conclusion}
In 5G/6G beam-based V2V communications operating at mmWave/sub-THz frequencies, the dynamic LOS link blockage due to obstructing vehicles is the dominant source of SNR loss, hindering the development of advanced safety-critical services in the context of autonomous driving. The vehicle-induced blockage depends on the number of potential vehicles blocking the LOS link as well as on their sizes. This paper proposes a novel analytical blockage modelling method, deriving a closed-form SNR distribution that takes into account the traffic density, the random 3D occupancy region of each vehicle, a minimum inter-vehicle safety distance, and separately accounts for the case in which TxV and RxV are located on the same lane or on different lanes. Moreover, as indicated by 3GPP, we account for both rooftop and bumper antenna placements. 
Numerical simulations validate the proposed analytical model, showing a good match between the analytical and the simulated SNR in all conditions. The presented analytical model is proved to be a tool for predicting the performance of V2V networks under blockage from dynamic objects such as surrounding vehicles. A significant service probability loss is observed when TxV and RxV are located on the same lane, and when the antenna arrays are mounted on the bumper, suggesting the usage of a rooftop-mounted antenna equipment. The results also show that V2V communication at mmW/sub-THz can only support short-range applications, while other solutions, e.g., relaying from other vehicles, are needed to increase the range without outages.


\bibliographystyle{IEEEtran}
\bibliography{reference}

\begin{thebibliography}{10}
\providecommand{\url}[1]{#1}
\csname url@samestyle\endcsname
\providecommand{\newblock}{\relax}
\providecommand{\bibinfo}[2]{#2}
\providecommand{\BIBentrySTDinterwordspacing}{\spaceskip=0pt\relax}
\providecommand{\BIBentryALTinterwordstretchfactor}{4}
\providecommand{\BIBentryALTinterwordspacing}{\spaceskip=\fontdimen2\font plus
\BIBentryALTinterwordstretchfactor\fontdimen3\font minus
  \fontdimen4\font\relax}
\providecommand{\BIBforeignlanguage}[2]{{%
\expandafter\ifx\csname l@#1\endcsname\relax
\typeout{** WARNING: IEEEtran.bst: No hyphenation pattern has been}%
\typeout{** loaded for the language `#1'. Using the pattern for}%
\typeout{** the default language instead.}%
\else
\language=\csname l@#1\endcsname
\fi
#2}}
\providecommand{\BIBdecl}{\relax}
\BIBdecl

\bibitem{3GPP2018}
3rd Generation Partnership Project~(3GPP), ``Study on enhancement of 3gpp
  support for 5g v2x services [tr 22.886],'' vol. Release 17, 2018.12.

\bibitem{coll2019sub}
B.~Coll-Perales, J.~Gozalvez, and M.~Gruteser, ``Sub-6ghz assisted mac for
  millimeter wave vehicular communications,'' \emph{IEEE Communications
  Magazine}, vol.~57, no.~3, pp. 125--131, 2019.

\bibitem{hakeem20205g}
S.~A.~A. Hakeem, A.~A. Hady, and H.~Kim, ``5g-v2x: Standardization,
  architecture, use cases, network-slicing, and edge-computing,''
  \emph{Wireless Networks}, vol.~26, no.~8, pp. 6015--6041, 2020.

\bibitem{jameel2018propagation}
F.~Jameel, S.~Wyne, S.~J. Nawaz, and Z.~Chang, ``Propagation channels for
  mmwave vehicular communications: State-of-the-art and future research
  directions,'' \emph{IEEE Wireless Communications}, vol.~26, no.~1, pp.
  144--150, 2018.

\bibitem{3GPPTR38901}
3rd Generation Partnership Project~(3GPP), ``Study on channel model for
  frequencies from 0.5 to 100 ghz,'' vol. TR 38.901, Tech. Rep., 2019.12.

\bibitem{boban2019multi}
M.~Boban, D.~Dupleich, N.~Iqbal, J.~Luo, C.~Schneider, R.~M{\"u}ller, Z.~Yu,
  D.~Steer, T.~J{\"a}ms{\"a}, J.~Li \emph{et~al.}, ``Multi-band
  vehicle-to-vehicle channel characterization in the presence of vehicle
  blockage,'' \emph{IEEE access}, vol.~7, pp. 9724--9735, 2019.

\bibitem{park2017millimeter}
J.-J. Park, J.~Lee, J.~Liang, K.-W. Kim, K.-c. Lee, and M.-D. Kim, ``Millimeter
  wave vehicular blockage characteristics based on 28 ghz measurements,'' in
  \emph{2017 IEEE 86th Vehicular Technology Conference (VTC-Fall)}.\hskip 1em
  plus 0.5em minus 0.4em\relax IEEE, 2017, pp. 1--5.

\bibitem{R1-1807672}
G.~T. R. W.~M. n.93, ``V2x sidelink channel model,'' vol. R1-1807672, May
  20178.

\bibitem{3GPPTR37885}
3rd Generation Partnership Project~(3GPP), ``Study on evaluation methodology of
  new vehicle-to-everything (v2x) use cases for lte and nr (v15.3.0, release
  15),'' vol. TR 37.885, Tech. Rep., 2019.06.

\bibitem{boban2016modeling}
M.~Boban, X.~Gong, and W.~Xu, ``Modeling the evolution of line-of-sight
  blockage for v2v channels,'' in \emph{2016 IEEE 84th Vehicular Technology
  Conference (VTC-Fall)}.\hskip 1em plus 0.5em minus 0.4em\relax IEEE, 2016,
  pp. 1--7.

\bibitem{giordani2019path}
M.~Giordani, T.~Shimizu, A.~Zanella, T.~Higuchi, O.~Altintas, and M.~Zorzi,
  ``Path loss models for v2v mmwave communication: performance evaluation and
  open challenges,'' in \emph{2019 IEEE 2nd Connected and Automated Vehicles
  Symposium (CAVS)}.\hskip 1em plus 0.5em minus 0.4em\relax IEEE, 2019, pp.
  1--5.

\bibitem{eshteiwi2020impact}
K.~Eshteiwi, G.~Kaddoum, B.~Selim, and F.~Gagnon, ``Impact of co-channel
  interference and vehicles as obstacles on full-duplex v2v cooperative
  wireless network,'' \emph{IEEE Transactions on Vehicular Technology},
  vol.~69, no.~7, pp. 7503--7517, 2020.

\bibitem{abul2007modeling}
A.~Abul-Magd, ``Modeling highway-traffic headway distributions using
  superstatistics,'' \emph{Physical Review E}, vol.~76, no.~5, p. 057101, 2007.

\bibitem{cui2018vehicle}
Q.~Cui, N.~Wang, and M.~Haenggi, ``Vehicle distributions in large and small
  cities: Spatial models and applications,'' \emph{IEEE Transactions on
  Vehicular Technology}, vol.~67, no.~11, pp. 10\,176--10\,189, 2018.

\bibitem{akhtar2014vehicle}
N.~Akhtar, S.~C. Ergen, and O.~Ozkasap, ``Vehicle mobility and communication
  channel models for realistic and efficient highway vanet simulation,''
  \emph{IEEE Transactions on Vehicular Technology}, vol.~64, no.~1, pp.
  248--262, 2014.

\bibitem{SavazziSpagnolini}
S.~Savazzi, S.~Sigg, M.~Nicoli, V.~Rampa, S.~Kianoush, and U.~Spagnolini,
  ``Device-free radio vision for assisted living: Leveraging wireless channel
  quality information for human sensing,'' \emph{IEEE Signal Processing
  Magazine}, vol.~33, no.~2, pp. 45--58, 2016.

\bibitem{GeoProb}
L.~Santalo, \emph{Integral Geometry and Geometric Probability}.\hskip 1em plus
  0.5em minus 0.4em\relax Encyclopedia of Mathematics and Its Applications,
  Addison-Wesley, 1976.

\bibitem{10.2307/3212475}
V.~S. Alagar, ``The distribution of the distance between random points,''
  \emph{Journal of Applied Probability}, vol.~13, no.~3, pp. 558--566, 1976.

\end{thebibliography}

\end{document}